\documentstyle{article}
\begin{document}
\noindent
{\bf Education Curricula of the Regional Centres for Space Science and Technology Education (Affiliated to the United Nations)}\par
\medskip
\noindent
Hans J. Haubold\par
\medskip
\noindent
Office for Outer Space Affairs, United Nations, Vienna International Centre, P.O. Box 500, 1400 Vienna, Austria (hans.haubold@unvienna.org)\par
\bigskip
\noindent
{\bf ABSTRACT:} Since 1988, the United Nations, through the Programme on Space Applications, is supporting the establishment and operation of regional Centres for Space Science and Technology Education in Africa, Asia and the Pacific, Latin America and the Caribbean, and Western Asia. Simultaneously, education curricula have been developed for remote sensing and geographic information system, satellite communications, satellite meteorology and global climate, and space and atmospheric science. The paper reviews briefly these developments and highlights the most recent updated education curricula in the four disciplines that are made available in 2002, in the six official languages of the United Nations, for implementation at the regional Centres and beyond.\par  
\medskip
Space science and technology education can be pursued at the elementary, secondary and university levels. In space faring nations, National Science Education Standards for space science and technology have been applied to science curricula at those levels (National Research Council, 1996). Such an innovation has not taken place in many developing nations, partly because the benefits of space science and technology have not been appreciated enough and partly because the facilities and resources for teaching science and technology at educational institutions are not yet well developed. Education in space science and technology in developed nations has become highly interactive; the World Wide Web and other information technologies have become useful tools in education programmes at all levels.\par
\medskip
   The incorporation of elements of space science and technology into university-level science curricula can serve a dual purpose for developed and developing nations (Daniel, 2002). It can enable nations to take advantage of the benefits inherent in the new technologies, which, in many cases, are spin-offs from space science and technology. It can also revitalize the educational system, introduce the concepts of high technology in a non-esoteric fashion and help create national capacities in science and technology in general. In that regard, Pyenson and Sheets-Pyenson (1999) emphasized in their recent work entitled 'Servants of Nature' that:\par
\medskip
{\it ``Both geographical decentralization and interdisciplinary innovation have become watchwords in academic science. Electronic information processing to some extent obviates the necessity for a scientist or scholar to reside at an ancient college of learning. Universities everywhere have adapted to new socioeconomic conditions by expanding curricula. They have always responded in this way, although never as quickly as their critics would like. Measured and deliberate innovation is one of academia's heavy burdens. It is also a great strength. Emerging fields of knowledge become new scientific disciplines only after they have found a secure place in universities. We look to universities for an authoritative word about the latest innovations. New scientific ideas emerge in a variety of settings, but they become the common heritage of humanity only when processed by an institution for advanced instruction like the modern university.''}\par
\medskip
   There are many challenges in the teaching of science at university level, both in developing and developed nations, but the challenges are of a higher magnitude in developing nations. The general problem confronting science education is the inability of students to see or experience the phenomena being taught, which often leads to an inability to learn basic principles and to see the relationship between two or more concepts and their practical relevance to problems in real life. Added to those problems is a lack of skills in the relevant aspects of mathematics and physics and in problem-solving strategies. There are also language problems in nations in which science is not taught in the national language(s). Over the years, developed nations have overcome most of the basic problems, except perhaps a psychological problem, namely that students may consider science to be a difficult subject. In developing nations, however, basic problems linger, exacerbated by the fact that there are not enough academically and professionally well-trained teachers.
 
   The General Assembly of the United Nations, in its resolution 45/72 of 11 December 1990, endorsed the recommendation of the Working Group of the Whole of the Scientific and Technical Subcommittee, as endorsed by the Committee on the Peaceful Uses of Outer Space, that the United Nations should lead, with the active support of its specialized agencies and other international organizations, an international effort to establish regional centres for space science and technology education in existing national/regional educational institutions in the developing nations.\footnote{UN Document A/AC.105/456, annex II, para. 4 (n), Report of the Scientific and Technical Subcommittee on the Work of its Twenty-Seventh Session, 12 March 1990, United Nations, New York.}\par
\medskip
   The General Assembly of the United Nations, in its resolution 50/27 of 6 December 1995, also endorsed the recommendation of the Committee on the Peaceful Uses of Outer Space that those centres be established on the basis of affiliation to the United Nations as early as possible and that such affiliation would provide the centres with the necessary recognition and would strengthen the possibilities of attracting donors and of establishing academic relationships with national and international space-related institutions.\footnote{UN Document A/RES/50/27, paragraph 30, International Cooperation in the Peaceful Uses of Outer Space, 5 February 1996, United Nations, New York.}\par
\medskip 
   Regional centres have been established in India for Asia and the Pacific\footnote{See web site of the regional Centre for Asia and the Pacific at http://www.cssteap.org/.}, in Morocco and Nigeria for Africa, in Brazil and Mexico for Latin America and the Caribbean, and in Jordan for Western Asia, under the auspices of the Programme on Space Applications, implemented by the Office for Outer Space Affairs.\footnote{UN Document A/AC.105/749, Regional Centres for Space Science and Technology Education (affiliated to the United Nations), 27 December 2000, United Nations, Vienna.}
 The objective of the centres is to enhance the capabilities of Member States, at the regional and international levels, in various disciplines of space science and technology that can advance their scientific, economic, and social development. Each of the centres provides postgraduate education, research, and application programmes with emphasis on (i) remote sensing and geographic information system, (ii) satellite communications, (iii) satellite meteorology and global climate, and (iv) space and atmospheric science for university educators and research and application scientists. Since 1996, all centres are implementing nine-month postgraduate courses in the four areas based on model curricula that emanated from the United Nations Meeting of Experts on the Development of Education Curricula for the Regional Centres for Space Science and Technology Education, held in Granada, Spain, in 1995. Since 1995, the curricula\footnote{UN Document A/AC.105/649, Centres for Space Science and Technology Education: Education Curricula, United Nations, Vienna; the document is also available in English and French at http://www.oosa.unvienna.org/SAP/centres/centres.htm.}
have been reviewed at regional and international educational meetings.\par
\medskip
   The Third United Nations Conference on the Exploration and Peaceful Uses of Outer Space (UNISPACE III, 1999), recommended that collaboration should be established between the regional centres and other national, regional and international organizations to strengthen components of their education curricula.\footnote{Report of the Third United Nations Conference on the Exploration and Peaceful Uses of Outer Space, Vienna, 9-30 July 1999 (United Nations publication, Sales No. E.00.I.3), chap. II, sect. G, para. 220; the document is also available at http://www.oosa.unvienna.org/.}
 In its resolution 54/68 of 6 December 1999, the General Assembly of the United Nations endorsed the resolution of UNISPACE III entitled "The Space Millennium: Vienna Declaration on Space and Human Development", in which action was recommended to ensure sustainable funding mechanisms for the regional centres.\footnote{Ibid., chap. I, resolution 1, para. 1 (e) (ii). The Declaration is also available on the home page of the Office for Outer Space Affairs (http://www.oosa.unvienna.org).}\par
\medskip
   The Office for Outer Space Affairs of the Secretariat of the United Nations organized, in cooperation with the European Space Agency, a second United Nations Expert Meeting on the Regional Centres for Space Science and Technology Education: Status and Future Development in Frascati, Italy, in 2001.\footnote{UN Document A/AC.105/782, Regional Centres for Space Science and Technology Education (affiliated to the United Nations), 14 March 2002, United Nations, Vienna.}
  The Meeting reviewed the status of establishment and operation of the regional centres with a view to enhancing cooperation between the centres. However, the main objective of the Meeting was to review and update the curricula at the university level and across cultures in the four areas. The Meeting considered that education varied significantly between nations and even between institutions within the same country which led to differences in space science and technology education curricula in terms of course content and modes of presentation. The Meeting noted that the model curricula\footnote{UN Document A/AC.105/649, Centres for Space Science and Technology Education: Education Curricula, United Nations, Vienna; the document is also available in English and French at http://www.oosa.unvienna.org/SAP/centres/centres.htm.}
had contributed to resolving such problems.\par
\medskip
   The Meeting updated the four education curricula and drew up course syllabuses that differ from most of those available in literature and on the World Wide Web. They are based on physics, mathematics and engineering as taught in many universities around the world. They are not tailored to any specific space-related project or mission that may have been or will be executed by any specific institution. The curricula are available, in the six official languages (Arabic, Chinese, English, French, Russian, Spanish) of the United Nations, on request, as United Nations documents.\footnote{UN Documents A/AC.105/L.238 (satellite meteorology and global climate), A/AC.105/L.239 (satellite communications), A/AC.105/L.240 (space and atmospheric science), and A/AC.105/L.241 (remote sensing and geographic information system).}\par
\bigskip
\noindent
{\bf LIST OF REFERENCES}\par
\medskip
\noindent
Daniel, R.R. (ed.) (2002). Concepts in Space Science, Universities Press, Orient Longman, Hyderabad, India.\par
\smallskip
\noindent
National Research Council (1996). National Science Education Standards, National Academy Press, Washington, D.C.\par
\smallskip
\noindent
Pyenson, L. and Sheets-Pyenson, S. (1999). Servants of Nature: A History of Scientific Institution, Enterprises, and Sensibilities, W. W. Norton and Company, New York.
\end{document}